\documentclass[sigconf, balance=false, natbib=true]{acmart}

\setcopyright{acmcopyright}
\copyrightyear{2024}
\acmYear{2024}
\acmConference[ISSAC '24]{ISSAC 2024}{July 16--19, 2024}{Raleigh, NC, USA}
\acmDOI{10.1145/3666000.3669695}

\pdfoutput=1
\usepackage{hyperref}    % hyperlinks
\usepackage{url}            % simple URL typesetting
\usepackage{booktabs}       % professional-quality 
\usepackage{algpseudocode,algorithm,algorithmicx}
\usepackage{cleveref}
\usepackage{tabularx}
\usepackage{adjustbox}

\usepackage{doi}
\usepackage{float}
\newtheorem{definition}{Definition}

\theoremstyle{definition}

\newcommand\numberthis{\addtocounter{equation}{1}\tag{\theequation}}

\begin{document}

\title{Faster Gr\"obner bases for Lie derivatives of ODE systems via monomial orderings}

\author{Mariya Bessonov}
\affiliation{%
    \institution{Department of Mathematics, CUNY NYC College of Technology}
    \city{New York}
    \state{NY}
    \country{USA}}
\email{mariya.bessonov@gmail.com}
\author{Ilia Ilmer}
\affiliation{%
    \institution{Ph.D. Program in Computer Science, CUNY Graduate Center}
    \city{New York}
    \state{NY}
    \country{USA}
}
\email{i.ilmer@icloud.com}
\author{Tatiana Konstantinova}
\affiliation{%
    \institution{Department of Mathematics, CUNY Queens College}
    \city{New York}
    \state{NY}
    \country{USA}}
\email{tatiana.v.konst@gmail.com}

\author{Alexey Ovchinnikov}
\affiliation{%
    \institution{Department of Mathematics, CUNY Queens College; Ph.D. Programs in Mathematics and Computer Science, CUNY Graduate Center}
    \city{New York}
    \state{NY}
    \country{USA}}
\email{aovchinnikov@qc.cuny.edu}

\author{Gleb Pogudin}
\affiliation{%
    \institution{LIX, CNRS, \'Ecole Polytechnique, Institute Polytechnique de Paris}
    \city{Paris}
    \country{France}
}
\email{gleb.pogudin@polytechnique.edu}

\author{Pedro Soto}\authornote{Work was partially done at the Department of Mathematics at Virginia Tech, the Mathematical Institute at the University of Oxford, and the Wellcome Centre for Human Genetics at the University of Oxford.}
\affiliation{%
    \institution{Ph.D. Program in Computer Science, CUNY Graduate Center}
    \city{New York}
    \state{NY}
    \country{USA}
}
\email{pedrosoto@vt.edu}

\renewcommand{\shortauthors}{Bessonov, et al.}

\begin{abstract}
Symbolic computation for systems of differential equations is often computationally expensive. Many practical differential models have a form of polynomial or rational ODE system with specified outputs. A basic symbolic approach to analyze these models is to compute and then symbolically process the polynomial system obtained by sufficiently many Lie derivatives of the output functions with respect to the vector field given by the ODE system.

In this paper, we present a method for speeding up  Gr\"obner basis computation for such a class of polynomial systems by using 
        specific monomial ordering, including weights for the variables, coming from the structure of the ODE model.
    We provide empirical results that show improvement across different symbolic computing frameworks and apply the method to speed up structural identifiability analysis of ODE models.
\end{abstract}


\maketitle

\section{Introduction}
Differential equations are widely used in modeling. 
Symbolic computation via differential algebra provides a broad range of tools for analyzing such models{ ~\cite{ollivier_hdr}}. 
However, efficiency has been a significant bottleneck in using such tools. 
There has been much progress in efficiency for ODE systems with specified output functions by symbolically processing the Lie derivatives of the output functions using Gr\"obner bases. 
However, for some particular examples of relatively small ODE systems (even $<10$ equations), the computation would not finish in weeks consuming over 100GB of RAM { (see, e.g.~\cite[Table~4]{dong2022differential} and~\cite[Table~6.1]{hong_global_2020})}.

The Gr\"obner basis (itself and its computation) of a polynomial system can vary based on the chosen monomial ordering.
The most common and empirically reliable in terms of computing time monomial ordering is the so-called total-degree-reverse-lexicographic order, or {\tt tdeg} in {\sc Maple} notation.
Weighted ordering adds a layer of comparison to monomial orderings where one first compares variables by the weight value multiplied by its degree exponent and then breaks ties by applying any applicable monomial rule{ ~\cite{faugere2016complexity}}.
Properly chosen weights may have tremendous impact on the computation time.
{  To illustrate this, c}onsider the following motivating example of a well-known benchmark polynomial system, Jason-210  \citep{eder2021efficient}. This example shows benefits of weights in general:
\begin{equation}\label{eq:motiv_example}
    P:=\begin{cases}
        x_1^2x_3^4 + x_1x_2x_3^2x_5^2 + x_1x_2x_3x_4x_5x_7 + x_1x_2x_3x_4x_6x_8 + \\
        + x_1x_2x_4^2x_6^2 + x_2^2x_4^4,                                          \ \
        x_2^6, \ \  ~x_1^6
    \end{cases}
\end{equation}
Computing the Gr\"obner basis of this system with {\tt tdeg}-order of \(x_1, x_2, x_3, x_4, x_5,\) \( x_6,x_7,x_8\) takes approximately \emph{670 seconds} of total CPU time and 26 {seconds} of total elapsed time {  (multiple cores were used)} as computed in Maple 2021\footnote{Computation done on MacBook Pro with 16 GB of RAM and 16-core M1 processor}.
Modifying the system by assigning a weight of 2 to the variable \(x_8\) results in approximately \emph{2 seconds} of CPU time and only about \emph{1 second} of total elapsed time. {  Assigning weights of $2$ to some of the other variables, e.g. to $x_7$, result in a speed-up as well.}

In this paper, we present a method for significantly speeding up Gr\"obner basis computation for the class of poylnomial systems that are formed by taking Lie derivatives and all variables and their derivatives are interpreted as indeterminates. Our method is based on a careful automatic selection of a monomial ordering, which is based on the structure of the ODE system. Our orderings are weighted  total-degree-reverse-lexicograpic, with the weight assignment following the ODE model structure. 

The presented ordering is a result of conducting numerous experiments and analyzing the results.
    It is thus
motivated by empirical observations, just like the fact that the total degree lexicographic ordering is also an empirical achievement accepted universally as the most advantageous monomial order.
Proving that the given choice behaves best is outside the scope of our manuscript, and we hope that our work will inspire more investigation in the area.
We provide experimental results showing improvements in runtime and memory use for {\sc Maple} and Magma.

One of the applied contexts in which such Lie derivative computation appears is the parameter identifiability problem.
 Parameter identifiability is a property crucial for designing high-quality mathematical models of real-world phenomena using ODEs. The question of  identifiability arises when one seeks a value for a particular parameter of the model.
A parameter can have either finitely many such values (\emph{local} structural identifiability), the value
can be unique (\emph{global} structural identifiability), or there may be infinitely many values and the parameter is \emph{unidentifiable}.

 Distinguishing between unidentifiable and locally identifiable is rather efficient~\cite{sedoglavic2002probabilistic}. On the other hand, knowing only local identifiability in practice is typically insufficient. For example, if one then uses an optimization-based parameter estimation algorithm~\cite{SBtoolbox,parameter_estimation,COPASI,EHCVMDBS2014,AMIGO2,Data2Dynamics}, one typically obtains only one solution for the parameter values even if there are multiple solutions fitting into a physically meaningful range. Knowing whether the system is globally identifiable would give the user a guarantee that the solution returned by the parameter estimation algorithm is unique.

The Gr\"obner basis computation with Lie derivatives described above, for instance, lies at the core of  the global identifiability algorithm SIAN~\citep{hong_sian_2019,ilmer_web-based_2021,ovchinnikov2020computing} { (see further details in Section~\ref{sec:sian_summary})}, used in~\citep{dankwa2021estimating,tran2020delicate,zhang2021integrated,alrdhawi2021longterm, locke2021quantification}.
We refer to a recent survey~\cite[Table~3]{ReyBarreiro2023} showing that SIAN compares favorably to other identifiability software tools.
The orderings proposed in the paper allow to speed up global identifiability analysis with SIAN significantly, and are included in the latest release of SIAN~\cite{hong_sian_2019} and SIAN-Julia \citep{sian_julia_github}.

The rest of this paper is organized as follows. In Section~\ref{relwork}, we provide an overview of works related to identifiability and Gr\"obner basis computation. Section~\ref{sec:preliminaries} describes Gr\"obner bases and how they appear in the identifiability analysis. Section~\ref{mainres} contains the weight generation algorithm. In Section~\ref{experiments}, we show the experimental results and benchmarks with our new weight assignment approach. We conclude in Section~\ref{conclusions} with final remarks regarding the work done and future directions of this research.

\section{Related work}
\label{relwork}

The analysis of connection between weights and homogenization of ideals appeared in~\citep{faugere2013complexity} and later in more detail in~\citep{faugere2016complexity}.
Homogeneous ideals are an intriguing special case of inputs for a Gr\"obner basis algorithm {  because of the additional structure~\cite[Section~10.2]{BWbook} and because it has been proven to lower the overall complexity of the F5 aglrotithm~\citep{faugere2013complexity, faugere2016complexity}}.
In the mentioned works, weights were used as a homogenization tool, { e.g., there are systems that can be homogenized by raising variables to the power given by a choice of weights.}
{  However, we} have observed in the motivating example~\eqref{eq:motiv_example} above that a weighted ordering can break homogenization, offering large benefits.

The problem of finding convenient variables orderings for Gr\"obner bases computation or similar tasks has recently been  actively investigated using  tools from machine learning~\citep{peifer2020learning,Florescu2020,England2020,Kauers2020}. 
These results typically allow arbitrary input systems and learn a black-box algorithm for choosing the ordering (for some recent work towards explainable ordering choice, see~\cite{Pickering2024}).
In this work, we focus on a specific class of input system only, but, for this class, we were able to find a simple human-understandable rule, which incorporates  domain-specific information.

\section{Preliminaries}
\label{sec:preliminaries}

\subsection{Gr\"obner bases}
We begin by defining \emph{monomial orderings} and \emph{Gr\"obner bases}.
\begin{definition}[Monomial Orderings]
    \label{def:monord}
    A monomial ordering $<$ of a polynomial ring is a total order on the set of monomials such that, for all monomials \(M_1,~M_2,~M_3\), we have: $1 \leq M_1$ and $M_1 < M_2\implies M_1M_3 < M_2M_3$.
\end{definition}
\begin{definition}[Gr\"obner Basis]
    \label{def:gb}
    Fix a monomial ordering \(<\) on the polynomial ring \(k[x_1,\dots,x_n]\). A 
    subset \(G=\{g_1,\dots,g_m\}\) of an ideal \(I\subseteq k[x_1,\dots,x_n]\) such that \(G\neq \{0\}\), is called \emph{Gr\"obner basis} if \[\langle LT(g_1),\dots,LT(g_m)\rangle = \langle LT(I)\rangle\]
    where \(LT(g_i)\) is the leading term of  \(g_i\), \(LT(I)\) are the leading terms of nonzero elements of \(I\), and \( \langle LT(I)\rangle\) is the ideal generated by \(LT(I)\).
\end{definition}

\subsection{Differential algebra and ODE systems with parameters}
In this section, we set up the language we will use to connect ODE systems and polynomial systems.
\begin{definition}[Differential rings and fields]
    A {\em differential ring} $(R,\delta)$ is a commutative ring with a derivation $\delta:R\to R$, that is, a map such that, for all $a,b\in R$, $\delta(a+b) = \delta(a) + \delta(b)$ and $\delta(ab) = \delta(a)b + a \delta(b)$.
    A differential ring that is also a field is called \emph{a differential field}.
\end{definition}

\begin{definition}[Differential polynomials and differential ideals]
    The {\em ring of differential polynomials} in the variables $x_1,\ldots,x_n$ over a field $K$ is the ring $K[x_j^{(i)}\mid i\geqslant 0,\, 1\leqslant j\leqslant n]$ with a derivation defined on the ring by \[\delta\left(x_j^{(i)}\right) := x_j^{(i + 1)}.\]
    This differential ring is denoted by $K\{x_1,\ldots,x_n\}$.
    An ideal $I$ of a differential ring $(R, \delta)$ is called a {\em differential ideal} if, for all $a \in I$, we have $\delta(a)\in I$. For $F\subset R$, the smallest differential ideal containing set $F$ is denoted by $[F]$.
\end{definition}

     For an ideal $I$ and element $a$ in a ring $R$, we denote \[I \colon a^\infty = \{r \in R \mid \exists \ell\colon a^\ell r \in I\}.\]
    This set is  an ideal in $R$.

\begin{definition}[Model in the state-space form]
    \normalfont{} A model in the \emph{state-space} form is a system
    \begin{equation}
        ~\Sigma :=~\begin{cases}
            \dot{\mathbf{x}} & =\mathbf{f}(\mathbf{x},~\boldsymbol\mu,~\mathbf{u}),~ \\
            \mathbf{y}       & =\mathbf{g}(\mathbf{x},~\boldsymbol\mu,~\mathbf{u}),~ \\
            \mathbf{x}(0)    & =\mathbf{x}^\ast,
        \end{cases}
        \label{eq:ode}
    \end{equation}
    where~\(\mathbf{f}=(f_1,\dots, f_n)\) and~\(\mathbf{g}=(g_1,\dots,g_m)\) with~\(f_i=f_i(\mathbf{x},~\boldsymbol\mu,~\mathbf{u})\), \(g_i=g_i(\mathbf{x},~\boldsymbol\mu,~\mathbf{u})\) are rational functions over the 
    complex numbers~\(\mathbb{C}\).
    The vector~\(\mathbf{x}=(x_1,\dots,x_n)\) represents the time-dependent state variables and~\(\dot{\mathbf{x}}\) represents the derivative.
    The vectors \(\mathbf{u}=(u_1,\dots,u_s)\), \(\mathbf{y}=(y_1,\dots,y_m)\), \(\boldsymbol\mu=(\mu_1,\dots,\mu_{\lambda})\),  and \(\mathbf{x}^\ast=(x_1^\ast,\dots,x_n^\ast)\) represent the input variables, output variables, parameters, and initial conditions, respectively.
    \label{def1}
\end{definition}

The analytic notion of identifiability~\cite[Definition~2.5]{hong_global_2020} is equivalent~(see \cite[Proposition~3.4]{hong_global_2020} and~\cite[Proposition~4.7]{OPT19}) to the following algebraic definition.

We write $\mathbf{f} = \frac{\mathbf{F}}{Q}$ and $\mathbf{g} = \frac{\mathbf{G}}{Q}$, where $\mathbf{F}$ and $\mathbf{G}$ are tuples from $\mathbb{C}(\boldsymbol{\mu})[\mathbf{x}, \mathbf{u}]$
and $Q$ is the common denominator of $\mathbf{f}$ and $\mathbf{g}$.
Define the following differential ideal, where we use $\dot{}$ in place of $\delta$.
\begin{equation}\label{eq:Isigma}
    I_{\Sigma} := [Q\dot x_1 - F_1, \ldots, Q\dot x_n - F_n, Qy_1 - G_1, \ldots, Qy_m - G_m] \colon Q^\infty,
\end{equation}
which is in $\mathbb{C}(\boldsymbol\mu)\{\mathbf{x}, \mathbf{y}, \mathbf{u}\}$.
Observe that every solution of~\eqref{eq:ode} is a solution of $I_\Sigma$.

\begin{definition}[Generic solution]\label{def:generic_solution}
    A tuple 
    $(\mathbf{x}_{s}, \mathbf{y}_{s}, \mathbf{u}_{s})$
    from a differential field $K \supset \mathbb{C}(\boldsymbol{\mu})$ is called
    \emph{a generic solution} of~\eqref{eq:ode} if, for every differential polynomial $P \in \mathbb{C}(\boldsymbol{\mu})\{\mathbf{x}, \mathbf{y}, \mathbf{u}\}$,
    \[
         P(\mathbf{x}_{s}, \mathbf{y}_{s}, \mathbf{u}_{s}) = 0 \iff P \in I_{\Sigma}.
    \]
\end{definition}

\begin{definition}[Identifiability]
    \label{def:id}
    Let \(\mathbb{C}(\boldsymbol{\mu})\) be a field of functions in \(\boldsymbol{\mu}\) with complex coefficients. A function (or parameter) \(h\in \mathbb{C}(\boldsymbol{\mu})\) is said to be identifiable (or globally identifiable) { in model~\eqref{eq:ode}} if, for every generic solution \((\mathbf{x}_s,\mathbf{y}_s,\mathbf{u}_s)\) of ODE \eqref{eq:ode}, it follows that \(h\in \mathbb{C}(\boldsymbol{\mu}, \mathbf{y}_s, \dot{\mathbf{y}}_s,\dots, \mathbf{u}_s, \dot{\mathbf{u}}_s,\dots)\). The function $h$ is said to be locally identifiable if \(h\) is algebraic over the field \(\mathbb{C}(\boldsymbol{\mu}, \mathbf{y}_s, \dot{\mathbf{y}}_s,\dots, \mathbf{u}_s, \dot{\mathbf{u}}_s,\dots)\).
\end{definition}

\subsection{Lie-derivative algorithm for parameter identifiability in ODE models}\label{sec:sian_summary}
We will now  summarize (following~\citep{hong_sian_2019}) how the class of polynomial systems we consider can be computed from ODE models~\eqref{eq:ode}:
\begin{enumerate}
    \item The original differential system~\eqref{eq:ode} is transformed into a polynomial system in the functions' derivatives and parameters through successive differentiation of the original equations. \label{construction}
    \item Random values are sampled for \(\mathbf{x}^*,\boldsymbol{\mu}\) and the derivatives of $u_i$'s.
    Then the corresponding values of the derivatives of $y_i$'s are computed.
    Finally, the values for $y_i$'s and $u_i$'s are plugged into the polynomial system.
    This corresponds to sampling a random input-output pair for the model.
    \end{enumerate}

After these steps, the polynomial system is, for instance, ready for an immediate use for the parameter identifiability problem for the ODE system using Gr\"obner bases.  
  In particular,   we then check whether the sampled values \(\mathbf{x}^*,\boldsymbol{\mu}\) are the only possible  solutions of the specialized polynomial system. If yes, the corresponding parameter is globally identifiable. 
   Due to random sampling, this algorithm may produce incorrect results, but the probability of correctness can be made arbitrarily close to $1$ by choosing an appropriate sampling range~\cite[Theorem~4.2]{hong_global_2020}.

The aforementioned Gr\"obner basis computation is typically the bottleneck in the identifiability analysis. 
We would like to emphasize two features of this computation:
\begin{itemize}
    \item The uniqueness of the value of a coordinate can easily be  checked using a G\"obner basis with respect to \emph{any ordering}.
    \item The resulting Gr\"obner basis is typically \emph{simple}, e.g., for a globally identifiable system, it defines a maximal ideal.
\end{itemize}

Notice that, for simplicity of presentation, from this point on, we do not separately discuss input variables \(\mathbf{u}\).

\subsection{Toy example}
In this section, we will show how a concrete (toy) ODE model is transformed into a polynomial system, which will further be a subject for Gr\"obner basis computation.
Consider the following ODE model in state-space form:
\begin{equation}
    \label{ex:sian-example}
    \Sigma=\begin{cases}
        \dot{x} = ax + c^2 \\
        y = x,             \\
        x(0) = x^{\ast}
    \end{cases}
\end{equation}
Using the Jacobian-based termination criterion~\cite[Theorem 3.16 and Proposition 3.20]{hong_sian_2019}, which is not relevant in the context of this paper, we will differentiate the first and the second equations one and two times, respectively.
As a result, the following polynomial system will be obtained:
\[
    E^t=\left[
    y-x^{\ast}, \dot{x}-ax^{\ast}-c^2, \dot{x}-\dot{y}, \ddot{x} - a\dot{x}, \ddot{x}-\ddot{y}, \dddot{x} - a\ddot{x}, \dddot{y}-\dddot{x}\right].
\] 
Here, the superscript $t$ stands for ``truncated'', which is the wording we use to represent the fact that some of the equations were differentiated a smaller number of times (but still sufficiently many) than one would naively do.

Then,  to restrict to a random output trajectory, 
we randomly sample values for \(x^{\ast}\) and \(a\), substitute them into \(E^t\) and solve the resulting system for $y, \dot{y}, \ddot{y}, \dddot{y}$ (the solution will be unique thanks to the triangular shape of the system).
We will denote this solution by \(\hat{Y}:=\left[\hat{y}_0,~\hat{y}_1,~\hat{y}_2\right]\).
Then we  substitute the solution into \(E^t\) obtaining \(\widehat{E}^t\). For a sample of \(a = 119791, x^{\ast} = 139697,c = 75091\), we have:
\begin{equation}\label{eq:Ethat-toy}\widehat{E}^t=\begin{cases}
        139697-x^{\ast},\dot{x}-ax^{\ast}-c^2,             \\
        -\dot{x}+22373101608, -a\dot{x}+\ddot{x},          \\
        -\ddot{x}+2680096214723928, \dddot{x} - a\ddot{x}, \\
        -\dddot{x} + 321051405657994059048
    \end{cases}
    \end{equation}
Then comes the \emph{key step} of symbolically processing this polynomial system to determine the property of the ODE system, global identifiability of the parameters. This is done by computing a Gr\"obner basis of~\eqref{eq:Ethat-toy} (the ordering does not matter). A parameter/initial condition of the ODE model is  globally identifiable if and only if, modulo the basis, it reduces to a constant.  For example, for~\eqref{eq:Ethat-toy}, we obtain a basis 
\[\mathcal{B}=\begin{cases}
        a-119791, -139697+x^{\ast},                     \\
        \dot{x}-22373101608, \ddot{x}-2680096214723928, \\
        \dddot{x} - 321051405657994059048, c^2 - 5638658281.
    \end{cases}\]
Notice that we have \(a-119791,~x^{\ast}-139697\) in the basis, so the reductions of $a$ and $x^\ast$ will be constants yielding that these parameters are globally identifiable. At the same time, we do not have a unique value for \(c\) thus concluding that it is \emph{only locally identifiable}.

\underline{Our goal}: find a weight assignment to each variable of the  polynomial system \(\widehat{E}^t\) to speed up the Gr\"obner basis computation. {In the example above, the weighted ordering would be applied before the step of computing the Gr\"obner basis, but after we generate a sampled system \(\widehat{E}^t\).}
We provide more technical details about how exactly this  is performed in \Cref{experiments}.

\section{Main result}
\label{mainres}

\subsection{The {monomial ordering.}}

{The monomial ordering we propose compares two monomials first by their weights (we describe the weight assignment below) and then breaks ties by 
  the reverse-lexicographic ordering in which  the variables are first compared with respect to their derivative order{ , e.g., $x< \dot x< \ddot x <\ldots$} (and then any ordering could be used, we used reverse alphabetical order, e.g., $\dot z > x > z$}).

Now we describe the key component of our monomial ordering, the \emph{weight assignment}.
Given a system~\eqref{eq:ode}, one can define \emph{the Lie derivative} $\mathcal{L}(h)$ of a function $h\in \mathbb{C}(\mathbf{x}, \boldsymbol{\mu}, \mathbf{u}, \mathbf{u}', \ldots)$ with respect to the system by
\begin{equation}
    \label{eq:lie}
    \mathcal{L}(h) = \sum\limits_{i=1}^{n} f_i \frac{\partial{h}}{\partial{x_i}}+\sum\limits_{j=1}^{s} \dot{u}_j\frac{\partial{h}}{\partial{u_j}}.
\end{equation}
By applying this formula to each output function \(y_i\), we can define, for each state variable or a parameter $a \in \{\mathbf{x}, \boldsymbol{\mu}\}$, 
the \emph{level} as
\begin{equation}
    \label{eq:level}
    \boxed{\mathrm{Level}(a) := \min\limits_{i} \left[ \exists \, y_j \in \mathbf{y} \colon a \text{ appears in }\mathcal{L}^i(y_j) \right]}.
\end{equation}
Using that value, we assign weight as follows:
\begin{itemize}
    \item for a state variable \(x_i\in x\) (and all its derivatives)
          \begin{equation}
              \label{eq:weight_states}
              \mathrm{Weight}(x_i) := \mathrm{Level}(x_i) + 1;
          \end{equation}
    \item for a parameter $\mu_i \in \boldsymbol{\mu}$:
          \begin{equation}
              \label{eq:weight_param}
              \mathrm{Weight}(\mu_i) := \begin{cases}
                  \mathrm{Level}(\mu_i) + 1, \text{ if } \mathrm{Level}(\mu_i)=\max\limits_{e\in \boldsymbol\mu\cup\mathbf{x}} \mathrm{Level}(e), \\
                  1, \text{ otherwise}.
              \end{cases}
          \end{equation}
\end{itemize}

\subsection{Example}
\label{sec:example}
Consider the following ODE system
\begin{equation}
    \label{ex:intro}
    \Sigma=\begin{cases}
        \dot{x}_1 = ax_1+bx_2, \\
        \dot{x}_2 = cx_1,      \\
        y_1 = x_1.
    \end{cases}
\end{equation}
Differentiating once:
\[
    \mathcal{L}(y_1) = \mathcal{L}(x_1) = ax_1+bx_2.
\]
We see that \(a, b, x_2\) all occur after the first differentiation and hence will have level of 1. At the same time, state \(x_1\) was already at level 0 and will not be considered further.
If we differentiate once more, we get
\[
    \mathcal{L}(\mathcal{L}(y_1)) = a\mathcal{L}(x_1)+b\mathcal{L}(x_2)= a(ax_1+bx_2)+ cx_2,
\]
bringing out \(c\). Differentiating further leads to no new information.  
The final weight assignment then is as follows: \[x_1\Rightarrow 1,\ \   x_2\Rightarrow 2\]

\subsection{{Do our weights homogenize the system?}}
We discussed earlier how it has been shown that polynomial systems benefit from homogenizing weight assignment (see \citep{faugere2013complexity} and \citep{faugere2016complexity}).  One may be tempted to hypothesize that homogenization would be the explanation behind the speed-up, but this does not happen because our systems are rarely homogeneous; we instead offer the hypothesis that avoiding reductions to zero, which we observed in our experiments, is the more likely cause of the speed-up, see Section~\ref{sec:explain}. Polynomial systems obtained by Lie derivatives in ODE models contain non-homogeneous polynomials in most cases by the nature of the problem statements and approach.
For example, consider an output function (see Definition~\ref{def1}) of the form
\begin{equation}
\label{eq:out}
    y_i=g_i(\dots),
\end{equation}
where $g_i$ is a polynomial.
Since polynomial elimination typically significantly speeds up after reducing the number of variables keeping the rest the same, the next  step we take is to replace the $y$-variables with numbers, such as in~\eqref{eq:Ethat-toy}.  
This way,~\eqref{eq:out} is inevitably transformed into an equation with a free term of degree 0.
Therefore, the polynomial systems from the class we consider  always have a non-homogeneous polynomial.

By design of our weight-assignment algorithm{ ~\eqref{eq:weight_param}}, the weight of any variable in \(g_i\) will be 1 { , since the variables of $g_i$ are exaxtly the base case of $i$ in Equation~\ref{eq:level}}. For other polynomials that do not have a free term and may be homogeneous, the maximum possible degree in the system will either increase or remain the same {  because we raise variables to the power of their weight similarly to the procedure described in \cite{faugere2016complexity}}. In this sense, we do not necessarily make polynomials ``more homogeneous''  with our weight assignment.

%%%%%%%%%%%%%%%%%%%%%%%%%%%%%%%%%%%%%%%%%%%%%%%%%%%%%%%%%

\subsection{Possible rationale behind the weight assignment}
\label{sec:explain}
{While the idea to use differential rev-lex ordering can be motivated by results in monomial ideals~\citep{Zobnin2009}, the mechanism behind the weight assignment seems to be more mysterious.}
In this section, we propose an explanation why the weight assignment \cref*{eq:weight_states,eq:weight_param} speeds up the computation.

We start with a brief overview of the F4 algorithm~\cite{faugere_new_1999}.
The original Buchberger algorithm~\citep{buchberger_theoretical_1976} iteratively picks a pair of polynomials $f, g$ from the already computed set and computes their S-polynomial
\begin{equation}
\label{eq:spoly}
\begin{gathered}    S(f, g) := \frac{M(f, g)}{\operatorname{LT}(f)}f - \frac{M(f, g)}{\operatorname{LT}(g)}g, \\ M(f, g) := \operatorname{lcm}(\operatorname{LM}(f), \operatorname{LM}(g)),
\end{gathered}
\end{equation}
where $\operatorname{LM}$ and $\operatorname{LT}$ stand for the leading monomial and leading term, respectively.
Then $S(f, g)$ is reduced with respect to already computed polynomials and the result, if nonzero, is added to the computed set.
The key idea of the F4 algorithm by~\cite{faugere_new_1999} is to select several S-polynomial at each step and then reduce them simultaneously using linear algebra.
This is done by constructing a matrix from the S-polynomials and all the multiples of the already computed polynomials which could be used in the reduction as follows: the columns correspond to the monomials appearing in at least one of the polynomials, so every polynomial can be transformed into a row in such a matrix.
We would like to point out two features of the algorithm important for our discussion:
\begin{itemize}
    \item The way a set of S-polynomials is chosen at each step may have dramatic impact on the performance of the algorithm.
          A popular approach is \emph{the normal strategy}~\cite[p. 73]{faugere_new_1999} which takes all pairs for which the \emph{formal degree}, $\deg M(f, g)$ (see~\eqref{eq:spoly}), is the minimal possible.
    \item The matrix is highly structured, in particular, the part containing the reducers (that is, not the S-polynomials) is by construction in a row echelon form and often has block-triangular shape.
          Therefore, the time for reducing such a matrix may depend more on the number of S-polynomials rather than the total number of rows in the matrix.

\end{itemize}
The S-polynomials which are reduced to zero can be considered as ``waste of time''.
\emph{Avoiding reductions} to zero is a recurring theme in the Gr\"obner bases computation, including the Buchberger criterion, F5 algorithm by~\cite{faugere_new_2002}, and connections to regular sequences~\cite[Section 2.4.3]{verron2016regularisation}.
We believe that one can explain the performance gains achieved by our weight assignment within this framework, although not directly through casting the system into a regular one.

Polynomial systems produced by our Lie derivative process for ODE systems typically have relatively small and simple Gr\"obner bases, so one may expect that few reductions are necessary.
On the other hand, the number of equations is large (starting with 40 in real examples) and the degrees are low (may not go beyond 3 in many applications).
Therefore, the normal selection strategy may select too many pairs at once, yielding a large number of zero reductions.
We claim that our weight assignment alleviates this issue by spreading possible values of $\deg M(f, g)$ used for selecting pairs.
Let us explain this in more detail using the following model as an example
\begin{align}
     & y = x_1,\label{eq:first_eq}                        \\
     & x_1' = -a x_1^2 + 2 b x_2,\label{eq:second_eq}     \\
     & x_2' = a x_1^2 - b x_2 - c x_2.\label{eq:third_eq}
\end{align}
This system corresponds to a chemical reaction network with two species $X_1$ and $X_2$ and reactions: \[2 X_1 \xrightarrow{a} X_2,\ X_2 \xrightarrow{b} 2X_1,\ X_2 \xrightarrow{c} \varnothing.\]
{  Variables $x_2$ and $c$ will be assigned a weight of 2 and 3, respectively and all the others will be assigned a weight of 1.}

By default, the algorithm will order variables ``alphabetically'' $x_1 > x_2$.
If we do not use weights, then the leading monomial of $\eqref{eq:first_eq}^{(i)}$, the $i$-th derivative of~\eqref{eq:first_eq}, will be $x_1^{(i)}$ while the leading monomials of~$\eqref{eq:second_eq}^{(i)}$ and~$\eqref{eq:third_eq}^{(i)}$ will be equal and come from $(ax_1^2)^{(i)}$  because these monomials will have higher total degree.
Now we consider the pairs of small formal degree.
In degree two, we will only have S-polynomials of derivatives of~\eqref{eq:first_eq}, which will be reducible to zero because the leading monomials in the pairs forming the S-polynomials are relatively prime (Buchberger's criterion).
Nontrivial pairs start with degree three, and there will be many of them, including
\begin{itemize}
    \item simple S-polynomials such as $S(\eqref{eq:first_eq}, \eqref{eq:second_eq})$, which basically correspond to plugging the known value for one of the $x_1$'s inside $x_1^2$ in~\eqref{eq:second_eq}{ , i.e., $S((13),(14)) = ax_1y + x_1'  - 2 bx_2$};
    \item less trivial S-polynomials such as $S(\eqref{eq:second_eq}^{(i)}, \eqref{eq:third_eq}^{(i)})$, which do not have such an immediate interpretation.
\end{itemize}

In the weighted case, the leading monomial of $\eqref{eq:first_eq}^{(i)}$ and $\eqref{eq:second_eq}^{(i)}$ will stay the same, while the leading monomial of $\eqref{eq:third_eq}^{(i)}$ will become $cx_2^{(i)}$.
This will change the situation significantly because the only remaining pairs of the  formal degree 3 { (earlier defined as $\deg M(f, g)$)} will be the natural ones corresponding to plugging the known values of $x_1$ and its derivatives to~\eqref{eq:second_eq} while considering more complicated S-polynomials is postponed.
As a result, the maximal number of pairs selected at a single iteration of F4 reduces from 20 in the no-weights case to 10 (see also Table~\ref{tab:F4stats}).

Generalizing this example, one can observe that the weight assignments attempt to force the variables of high level to appear in the leading monomials of the corresponding equations, thus making the pairs involving these equations to be of higher formal degree.
This will avoid selecting too many pairs at once and steer the computation towards first working with the variables and equations of small level only, thus taking advantage of the known $y$-values.
 We suspect this heuristic is particularly well-suited for ODEs in the form of Eq.~\ref{eq:ode} since the outputs are truly modeling ``known'' output values from real world applications. 
In particular, both globally identifiable and parameter estimation algorithms set the output functions equal to a number, and the heuristic for system solving would be to solve for unknown quantities that are ``close'' to the known quantities first;
our weight assignments precisely measure this ``closeness''.

We used \texttt{msolve}~\citep{msolve} to check whether using weights indeed reduces the maximal number of pairs selected at the same time and the number of zero reductions.
Thanks to being open source, \texttt{msolve} allows us to extract all information of interest easily.
The results are given in Table~\ref{tab:F4stats} and confirm our expectations.
We also double-checked smaller examples using our own basic implementation of F4 in {\sc Maple}\footnote{available at \url{https://github.com/iliailmer/BasicF4Algorithm}}, and observed the same phenomena.

\begin{table}[!ht]
    \centering
    \begin{tabular}{|l|c|c|c|c|}
        \hline
                                                        &
        \multicolumn{2}{|c|}{Max. \# of pairs selected} & \multicolumn{2}{|c|}{\# of zero reductions}                                  \\
        \hline
        Model                                           & No weights                                  & Weights & No weights & Weights \\
        \hline\hline
        \eqref{eq:first_eq}-\eqref{eq:third_eq}         & $20$                                        & $10$    & $21$       & $19$    \\
        \hline
        \eqref{seirp}                                     & $25451$                                     & $3472$  & $42857$    & $20581$ \\
        \hline

        \eqref{seir}                                      & $34570$                                     & $2731$  & $59804$    & $11546$ \\
        \hline
        \eqref{seir2}                                     & $10370$                                     & $2021$  & $27972$    & $8953$  \\
        \hline
        \eqref{nfkb}                                      & $10555$                                     & $6653$  & $27795$    & $18102$ \\
        \hline
    \end{tabular}
    \caption{F4 statistics on benchmarks with/without weights}
    \label{tab:F4stats}
\end{table}

%%%%%%%%%%%%%%%%%%%%%%%%%%%%%%%%%%%%%%%%%%%%%%%%%%%%%%%%%

\section{Experimental results}
\label{experiments}
In this section, we present several examples of ODE systems, for which we observe reduction in both the runtime and memory. All simulations were run on a cluster with 64 Intel Xeon CPU with 2.30GHz clock frequency and 755 GB RAM.\@ We ran the computation using {\sc Maple} and Magma computer algebra systems.

 The original SIAN algorithm~\cite{hong_global_2020} computed Gr\"obner bases over rationals.
However, many popular F4 implementations (including the one in {\sc Maple}) are multimodular, that is, the actual computation is in fact performed modulo several prime numbers, and then the result is lifted to the rationals. 
To reduce the uncertainty related to different possibilities of the choices of primes and focus on the performance gains of the proposed weighted ordering, we run all  experiments modulo a fixed prime number \(p=11863279\).
We have conducted additional experiments  to verify that the speedup is similar for the computations over rationals.

{\sc Maple} does not directly support the use of weighted orderings with a compiled F4 implementation that is sufficiently fast.
To avoid any potential slowdowns, we substitute any variable $v$ in the polynomial system that has weight $w$ greater than 1 with $v^w$.
To illustrate this, if we have a polynomial system \(E=\{x+y, x-y\},\)
and we wish to use the weight of \(2\) for variable \(x\), our approach is to compute the basis for a new polynomial system \(E_1=\{x^2+y, x^2-y\}\)
keeping the variable order as total degree reverse lexicographic.
Empirically, there may be a difference observed between computing Gr\"obner basis with \(x>y\) and \(y>x\). In our computations, we order the variables by the degree of the derivative. For example, consider a simple input ODE model \begin{equation}
    \begin{cases}
        \dot{x}_1 = ax_1, \\ \dot{x}_2 = -bx_1 + cx_2, \\
        y=x_1+x_2.
    \end{cases}
\end{equation}
We then produce the following polynomial system, where the double index in \(x_{i;j}\) shows that the variable is the \(j\)-th derivative of \(x_i\) in jet-notation.
\begin{equation}
    E=\begin{cases}
        7828371-x_{1;0}-x_{2;0}, -ax_1+x_{1;1},             \\
        bx_1-cx_2+x_{2;1}, -x_{1;1}-x_{2;1}+22382588610034, \\
        -ax_{1;1}+x_{1;1}, bx_{1;1}-cx_{2;1}+x_{2;2},       \\
        -x_{1;1}-x_{2;2}+98741152216384012556,              \\
        -ax_{1;1}+x_{1;3}, bx_{1;1}-cx_{2;2}+x_{2;3},       \\
        -x_{1;3}-x_{2;3}+538005180363000517510923144,       \\
        -ax_{1;3}+x_{1;4}, bx_{1;3}-
        cx_{2;3}+x_{2;4},                                   \\
        -x_{1;4}-x_{2;4}+3127015821351630984063385030338736
    \end{cases}
\end{equation}
the order of variables for the best speed without weights is
\begin{equation}
    \label{eq:default_order}
    x_{2;4}, x_{1;4}, x_{2;3}, x_{1;3}, x_{2;2}, x_{1;2}, x_{2;1}, x_{1;1}, x_{2;0}, x_{1;0}, a, b, c.
\end{equation}
That is, we use differential deg-rev-lex ordering which  orders variables from higher to lower derivative 
grouping the same degree together (all order-4 derivatives, 
all order-3, etc.).

In what follows, we apply the weights on top of the default variable ordering~\eqref{eq:default_order} that has proven itself to be empirically faster. We will consider several ODE models and provide Gr\"obner basis results over a field of integers with positive prime characteristic \(p=11863279\).
Each example will be summarized by the following metrics in Tables~\ref{benchmarks_maple} and~\ref{benchmarks_magma}:
\begin{enumerate}
    \item Number of polynomials and variables in the polynomial system.
          
    \item {Default (without weights) CPU time (min) and memory (GB)}.
         
    \item {CPU time (min) and memory (GB) with weights}.
          
    \item Speedup calculated as \(\frac{\text{old time}}{\text{new time}}\).
    \item Memory improvement calculated as \(\frac{\text{old memory}}{\text{new memory}}\).
\end{enumerate}
Once the Gr\"obner basis computation is finished, the weights are removed by a back substitution to answer the identifiability query.

\setlength{\tabcolsep}{1pt}
\renewcommand{\arraystretch}{0.7}

\begin{table}[!ht]
    {\tiny
        \begin{tabular}{||c|c|c||c|c|c|c||c|c|c|c||}
            \hline
            \multicolumn{3}{||c||}{Model information} & \multicolumn{4}{c||}{Time (min)} & \multicolumn{4}{c||}{Memory (GB)}                                                                                                       \\
            \hline
            Model                                     & num.                             & num.                              & old SIAN & differential & our final & speedup    & old SIAN & differential & our final & reduction  \\
            name                                      & polys.                           & vars.                             & ordering & degrevlex    & ordering  &            & ordering & degrevlex    & ordering  &            \\\hline\hline
            
            COVID Model 2,                            &                                  &                                   &          &              &           &            &          &              &           &            \\
            \eqref{ssaair}                              & 49                               & 48                                & N/A      & N/A          & 602.0     & \(\infty\) & N/A      & N/A          & 23.2      & \(\infty\) \\\hline
            Pharmacokinetics,                         &                                  &                                   &          &              &           &            &          &              &           &            \\
            \eqref{pharm}                               & 48                               & 47                                & N/A      & N/A          & 21.0      & \(\infty\) & N/A      & N/A          & 7.7       & \(\infty\) \\\hline
            HPV,                                      &                                  &                                   &          &              &           &            &          &              &           &            \\
            \eqref{HPV}, \eqref{hpv_gr1}                  & 97                               & 92                                & N/A      & N/A          & 13.9      & \(\infty\) & N/A      & N/A          & 3.7       & \(\infty\) \\\hline
            HPV,                                      &                                  &                                   &          &              &           &            &          &              &           &            \\
            \eqref{HPV}, \eqref{hpv_gr4}                  & 79                               & 75                                & N/A      & N/A          & 5.1       & \(\infty\) & N/A      & N/A          & 11.0      & \(\infty\) \\\hline
            COVID Model 1,                            &                                  &                                   &          &              &           &            &          &              &           &            \\
            \eqref{seiqrdc}                             & 51                               & 50                                & 377.0    & 321.9        & 1.0       & 327.6      & 15.3     & 15.2         & 0.3       & 52.6       \\\hline
            Goodwin Oscillator                        &                                  &                                   &          &              &           &            &          &              &           &            \\
            \eqref{goodwin}                             & 42                               & 43                                & 44.1     & 29.8         & 1.5       & 18.9       & 10.8     & 10.6         & 0.7       & 14.6       \\\hline
            SEIR-1,                                   &                                  &                                   &          &              &           &            &          &              &           &            \\
            \eqref{seir}                                & 44                               & 45                                & 3.5      & 2.2          & 0.1       & 17.4       & 3.3      & 3.3          & 0.1       & 44.8       \\\hline
            NF-\(\kappa\)B,                           &                                  &                                   &          &              &           &            &          &              &           &            \\
            \eqref{nfkb}                                & 120                              & 109                               & 10.6     & 7.1          & 2.3       & 3.0        & 11.8     & 6.1          & 3.1       & 1.9        \\\hline
            SEIRP,                                    &                                  &                                   &          &              &           &            &          &              &           &            \\
            \eqref{seirp}                               & 50                               & 42                                & 2.6      & 2.0          & 0.8       & 2.5        & 1.0      & 1.6          & 0.2       & 8.5        \\\hline
            SEIR-2,                                   &                                  &                                   &          &              &           &            &          &              &           &            \\
            \eqref{seir2}                               & 44                               & 43                                & 1.3      & 0.8          & 0.4       & 2.2        & 0.8      & 0.7          & 0.1       & 6.1        \\\hline
        \end{tabular}
    }
    \caption{Results of applying the weighted ordering to only Gr\"obner basis computation step of Lie derivative processing (SIAN algorithm), with characteristic \(p=11863279\) using {\sc Maple} 2021.2. We compare three monomial orderings: originally used in SIAN, differential degrevlex, and our main weighted ordering.
        ``N/A'' stands for the {\sc Maple} error
        ``\texttt{Error, (in Groebner:-F4:-GroebnerBasis) numeric exception:~division by zero}''   without clear direct cause.}
    \label{benchmarks_maple}
\end{table}

\begin{table}[!ht]
   {\tiny
        \begin{tabular}{||c|c|c||c|c|c|c||c|c|c|c||}
            \hline
            \multicolumn{3}{||c||}{Model information} & \multicolumn{4}{c||}{Time (min)} & \multicolumn{4}{c||}{Memory (GB)}                                                                                                   \\
            \hline
            Model                                     & num.                             & num.                              & old SIAN & differential & our final & speedup & old SIAN & differential & our final & reduction \\
            name                                      & polys.                           & vars.                             & ordering & degrevlex    & ordering  &         & ordering & degrevlex    & ordering  &           \\\hline\hline
            COVID Model 2,                            &                                  &                                   &          &              &           &         &          &              &           &           \\
            \eqref{ssaair}                              & 49                               & 48                                & 4000.6   & 3471.2       & 517.4     & 6.7     & 38.6     & 36.4         & 21.6      & 1.7       \\\hline
            Pharmacokinetics,                         &                                  &                                   &          &              &           &         &          &              &           &           \\
            \eqref{pharm}                               & 48                               & 47                                & 757.6    & 248.3        & 44.5      & 5.6     & 14.7     & 8.4          & 10.7      & 0.8       \\\hline
            HPV,                                      &                                  &                                   &          &              &           &         &          &              &           &           \\
            \eqref{HPV}, \eqref{hpv_gr1}                  & 97                               & 92                                & 321.7    & 126.6        & 51.5      & 2.4     & 21.4     & 9.8          & 18.6      & 0.5       \\\hline
            HPV,                                      &                                  &                                   &          &              &           &         &          &              &           &           \\
            \eqref{HPV}, \eqref{hpv_gr4}                  & 79                               & 75                                & 6.8      & 5.9          & 3.2       & 1.4     & 2.7      & 2.6          & 2.4       & 1.1       \\\hline
            COVID Model 1,                            &                                  &                                   &          &              &           &         &          &              &           &           \\
            \eqref{seiqrdc}                             & 51                               & 50                                & 1331.1   & 1272.1       & 0.6       & 2207.9  & 9.2      & 8.7          & 1.8       & 4.8       \\\hline
            Goodwin Oscillator                        &                                  &                                   &          &              &           &         &          &              &           &           \\
            \eqref{goodwin}                             & 42                               & 43                                & 26.9     & 22.4         & 0.8       & 28.5    & 3.5      & 3.1          & 0.5       & 6.0       \\\hline
            SEIR-1,                                   &                                  &                                   &          &              &           &         &          &              &           &           \\
            \eqref{seir}                                & 44                               & 45                                & 8.6      & 3.9          & 0.1       & 76.0    & 2.0      & 2.0          & 0.2       & 9.8       \\\hline
            NF-\(\kappa\)B,                           &                                  &                                   &          &              &           &         &          &              &           &           \\
            \eqref{nfkb}                                & 120                              & 109                               & 14.6     & 9.1          & 1.7       & 5.2     & 3.3      & 2.0          & 0.6       & 3.5       \\\hline
            SEIRP,                                    &                                  &                                   &          &              &           &         &          &              &           &           \\
            \eqref{seirp}                               & 50                               & 42                                & 10.0     & 6.8          & 36.5      & 11.2    & 1.4      & 1.3          & 1.6       & 0.8       \\\hline
            SEIR-2,                                   &                                  &                                   &          &              &           &         &          &              &           &           \\
            \eqref{seir2}                               & 44                               & 43                                & 3.4      & 1.2          & 0.2       & 7.6     & 2.0      & 1.2          & 0.7       & 1.6       \\\hline
        \end{tabular}
    }
    \caption{Results of applying the weighted ordering to only Gr\"obner basis computation step of Lie derivative processing (SIAN algorithm) with  characteristic \(p=11863279\) in Magma 2.26-8. We compare three underlying
            monomial orderings: originally used in SIAN, differential degrevlex, and our main weighted ordering.} 
    \label{benchmarks_magma}
\end{table}

\section{Concluding remarks}
\label{conclusions}

We presented an approach to automatically choose a {weighted monomial ordering}
for Gr\"obner basis computation for a class of polynomial systems obtained by computing Lie derivatives of output functions in ODE models. This is, for example, a key component of assessing parameter identifiability of the ODE models~\citep{hong_sian_2019,hong_global_2020}.
We observe significant improvements for multiple models that vary in complexity, number of polynomials, and number of variables.

Our main idea for weight generation lies in the observation that the ``closedness'' of
parameters and states in the ODE  to the outputs makes a difference for the effect of a weighted ordering. 
These empirical observations translated into a sequential Lie differentiation of output functions.
Effectively, this differentiation produces Taylor coefficients of output functions \(\mathbf{y}\) in terms of states at a fixed time \(t=0\) and parameters. We assign weights depending on the depth of these Taylor coefficients, thus, effectively, leveraging the outputs ``sensitivity''.

If the systems were already relatively quick to return the answer, the weights did not have a negative impact. In fact, in examples where computation slowed down {(see e.g. Section~\ref{sec:slowdown})}, the memory usage still showed a positive effect, decreasing by around 80\%. There was also a case where the program ran around 44\% faster but consumed 30\% more memory. These non-trivial examples constitute a minority of systems. In some cases, a user would not require a weighted ordering because the Gr\"obner basis computation runs {fast} without weights.

\begin{acks}
The authors are grateful to CCiS at CUNY Queens College for the computational resources, to Andrew Brouwer for pointing out the HPV example, to Mohab Safey El Din for bringing our attention to regular sequences, to Alexander Demin for helpful discussions about F4 algorithm, and to the referees for their useful comments. This work was partially supported by the NSF  grants CCF-2212460, 1563942, 1564132 and DMS-1760448, 1853650, 1853482, and the French ANR-22-CE48-0008 OCCAM project.
\end{acks}

\bibliographystyle{ACM-Reference-Format}
\bibliography{bibliography}

\section{Systems and weights}

In this section, we present details about models considered. Specifically, we will describe the differential equations and the resulting weights of the models used in the analysis of this paper.

\subsection{Goodwin oscillator}
This model is presented in~\eqref{goodwin} and comes from~\citep{goodwin_oscillatory_1965} and describes time periodicity in cell behavior. This example has 4 state variables \(x_{1,2,3,4}\) and 6 parameters. Below is the Goodwin oscillator model and the weight assignment for it (identity weights are not displayed):
\begin{align}
        \begin{cases}
                 \dot{x}_1 = -b  \, x_1 + \frac{1}{(c + x_4)},                                 \\
                 \dot{x}_2 = \alpha  \, x_1 - \beta  \, x_2,                                   \\
                 \dot{x}_3 = \gamma  \, x_2 - \delta  \, x_3,                                  \\
                 \dot{x}_4 = \frac{\sigma  \, x_4  \, (\gamma  \, x_2 - \delta  \, x_3)}{x_3}, \\
                 y = x_1
             \end{cases}
         &
        \begin{tabular}{lcl}
            $x_2$   & $\implies$ & 3 \\
            $x_3$   & $\implies$ & 3 \\
            $x_4$   & $\implies$ & 2 \\
            $\beta$ & $\implies$ & 4
        \end{tabular}
\label{goodwin}
\end{align}

SIAN uses an auxiliary variable \(z_{aux}\) to account for the presence of denominators in the right-hand side of the original input ODE system. We observe that giving a weight of at most 3 to this variable does not decrease performance.

\subsection{SEIRP model}

This is a biomedical model applied to COVID-19 in~\citep{massonis_structural_2020}.
The outputs were changed to make the system more of a computational challenge to SIAN. Below is the SEIRP model and the weight assignment for it (identity weights are not displayed):
\begin{align}
        \begin{cases}
                 \dot{S} = - \alpha_e \,S \,E - \alpha_i \,S \,I,                       \\
                 \dot{E} = \alpha_e \,S \,E + \alpha_i \,S \,I - \kappa \,E - \rho \,E, \\
                 \dot{I} = \kappa \,E - \beta \,I - \mu \,I,                            \\
                 \dot{R} = \beta \,I + \rho \,E,                                        \\
                 \dot{P} = \mu \,I,                                                     \\
                 y_1 = I + S
             \end{cases}
         &
        \begin{tabular}{lcl}
            $E$    & $\implies$ & 2 \\
            $\rho$ & $\implies$ & 3
        \end{tabular}
    \label{seirp}
\end{align}

\subsection{SEIR COVID-19 model}
Next we consider a SEIR-model of epidemics from~\cite[table 2, ID=14]{massonis_structural_2020}. The example originally had 3 output functions. We reduced it to 1 to create more of a computational challenge for our program. We also use the term \(\mu \,i \,s\) instead of \(\mu \,i\mu \,s\) in the third equation. The state-space form of the model and the weight assignment are presented in~\eqref{seiqrdc}:
\begin{align}
 \label{seiqrdc}
\begin{cases}
                 \dot{S} = \mu  \, N - \alpha  \, S - \beta  \, S  \, I  \, N - \mu  \, S, \\
                 \dot{E} = \beta  \, S  \, I  \, N - \mu  \, E - \gamma  \, E,             \\
                 \dot{I} = \gamma  \, E - \delta  \, I - \mu  \, I  \, S,                  \\
                 \dot{Q} = \delta  \, I - \lambda  \, Q - \kappa  \, Q - \mu  \, Q,        \\
                 \dot{R} = \lambda  \, Q - \mu  \, S,                                      \\
                 \dot{D} = \kappa  \, Q,                                                   \\
                 \dot{C} = \alpha  \, S - \mu  \, C - \tau  \, C ,                         \\
                 y = C
             \end{cases}
         &
        \begin{tabular}{lcl}
            $S$      & $\implies$ & 2 \\
            $I$      & $\implies$ & 3 \\
            $E$      & $\implies$ & 4 \\
            $\gamma$ & $\implies$ & 4 \\
            $\delta$ & $\implies$ & 4
        \end{tabular}
\end{align}

\subsection{SIR model with forcing term}
The following model was presented in~\citep{capistran_parameter_2009}.
This is a SIR-model with an oscillating forcing term given by equations
for \(x_1,~x_2\). We also give our weight assignment.
\begin{align}\begin{cases}                 \dot{S} = \mu - \mu \, S - b_0 \, (1 + b_1 \, x_1) \, I \, S + g \, R, \\
                 \dot{I} = b_0 \, (1 + b_1 \, x_1) \, I \, S - (\nu + \mu) \, I,        \\
                 \dot{R} = \nu \, I - (\mu + g) \, R,                                   \\
                 \dot{x}_1 = -M \, x_2,                                                 \\
                 \dot{x}_2 = M \, x_1,                                                  \\
                 y_1 = I,                                                               \
                 y_2 = R.
             \end{cases}
         &
        \begin{tabular}{lcl}
            $x_1$ & $\implies$ & 2 \\
            $x_2$ & $\implies$ & 3 \\
            $M$   & $\implies$ & 3
        \end{tabular}
\label{sirsforced}
\end{align}

\subsection{A different SEIR-like COVID-19 model}

The following model also comes from~\citep{massonis_structural_2020}. We also provide our weight assignment.
\begin{gather*}
        \begin{cases}
                 \dot{S}_d =  -\epsilon_s \,\beta_a \,(A_n + \epsilon_a \,A_d) \,S_d - h_1 \,S_d + h_2 \,S_n - \epsilon_s \,\beta_i \,S_d \,I_n, \\
                 \dot{S}_n = -\beta_i \,S_n \,I_n - \beta_a \,(A_n + \epsilon_a \,A_d) \,S_n + h_1 \,S_d - h_2 \,S_n,                            \\
                 \dot{A}_d = \epsilon_s \beta_i S_d I_n + \epsilon_s \beta_a (A_n + \epsilon_a A_d) S_n + h_2 A_n - \gamma_{ai} A_d - h_1 A_d,   \\
                 \dot{A}_n = \beta_i \,S_n \,I_n + \beta_a \,(A_n + \epsilon_a \,A_d) \,S_n + h_1 \,A_d - \gamma_{ai} \,A_n - h_2 \,A_n,         \\
                 \dot{I}_n = f \,\gamma_{ai} \,(A_d + A_n) - \delta \,I_n - \gamma_{ir} \,I_n,                                                   \\
                 \dot{R}   = (1-f) \,\gamma_{ai} \,(A_d + A_n) + \gamma_{ir} \,I_n,                                                              \\
                 y_1  = S_d,                                                                                                                     \ \ y_2  = I_n
             \end{cases}
         \\
        \begin{tabular}{lcl}
            $A_d$             & $\implies$ & 2 \\
            $A_n$             & $\implies$ & 2 \\
            $S_n$             & $\implies$ & 2 \\
            $\beta_{a, i}$    & $\implies$ & 2 \\
            $h_{1, 2}$        & $\implies$ & 2 \\
            $\gamma_{ai}$     & $\implies$ & 2 \\
            $f$               & $\implies$ & 2 \\
            $\epsilon_{a, s}$ & $\implies$ & 2
        \end{tabular}
        \label{ssaair}
    \end{gather*}
In this model, the computation without weights has not finished in reasonable time, consuming all available memory. 

\subsection{Pharmacokinetics model}
This model comes from~\citep{demignot_effect_1987} describing pharmacokinetics of glucose-oxidase. We make one modification setting \(a_1=a_2\). The model is small but presents a significant computational challenge for global identifiability,
that is, it is very difficult to compute Gr\"obner basis of this model's polynomial system in SIAN. We also provide our weight assignment:
\begin{align}
\begin{cases}
                 \dot{x}_1 = a_1 \,(x_2 - x_1) - \frac{(k_a \,n \,x_1)}{(k_c \,k_a + k_c \,x_3 + k_a \,x_1)}, \\
                 \dot{x}_2 = a_1 \,(x_1 - x_2),                                                               \\
                 \dot{x}_3 = b_1 \,(x_4 - x_3) - \frac{(k_c \,n \,x_3)}{(k_c \,k_a + k_c \,x_3 + k_a \,x_1)}, \\
                 \dot{x}_4 = b_2 \,(x_3 - x_4),                                                               \\
                 y_1 = x_1
             \end{cases}
         &
        \begin{tabular}{lcl}
            $x_2$ & $\implies$ & 2 \\
            $x_3$ & $\implies$ & 3 \\
            $x_4$ & $\implies$ & 3 \\
            $b_2$ & $\implies$ & 4
        \end{tabular}
\label{pharm}
\end{align}

\subsection{NF-\(\kappa\)B model}

This model comes from~\citep{tomasz_mathematical_2004} and was used for identifiability analysis in~\citep{balsa_iterative_2010}. The ODE system consists of 15 equations,~\eqref{nfkb},
\begin{equation}
    \label{nfkb}
    \begin{cases}
        \dot{x}_{1} =   k_{p} - k_{d} \, x_1 - k_1 \, x_1  \, u,                                             \\
        \dot{x}_{2} =   -k_3 \, x_2 - k_{d} \, x_2 - a_2 \, x_2 \, x_{10} +                                  \\
        + t_1 \, x_4 - a_3 \, x_2  \, x_{13} + t_2 \, x_5 + (k_1 \, x_1 - k_2 \, x_2  \, x_8) \, u,          \\
        \dot{x}_{3} =   k_3 \, x_2 - k_{d} \, x_3 + k_2 \, x_2  \, x_8  \, u,                                \\
        \dot{x}_{4} =   a_2 \, x_2  \, x_{10} - t_1 \, x_4,                                                  \\
        \dot{x}_{5} =  a_3 \, x_2  \, x_{13} - t_2 \, x_5,                                                   \\
        \dot{x}_{6} =  c_{6a} \, x_{13} - a_1 \, x_6  \, x_{10} + t_2 \, x_5 - i_1 \, x_6,                   \\
        \dot{x}_{7} =  i_1 \, k_v \, x_6 - a_1 \, x_{11}  \, x_7,                                            \\
        \dot{x}_{8} =  c_4 \, x_9 - c_5 \, x_8,                                                              \\
        \dot{x}_{9} =  c_2 + c_1 \, x_7 - c_3 \, x_9,                                                        \\
        \dot{x}_{10} = -a_2 \, x_2  \, x_{10} - a_1 \, x_{10} \, x_6 + c_{4a}  x_{12} -                      \\
        - c_{5a}  x_{10} - i_{1a}  x_{10} + e_{1a}  x_{11},                                                  \\
        \dot{x}_{11} =  -a_1 \, x_{11}  \, x_7 + i_{1a} \, k_v \, x_{10} - e_{1a} \, k_v \, x_{11},          \\
        \dot{x}_{12} =  c_{2a} + c_{1a} \, x_7 - c_{3a} \, x_{12},                                           \\
        \dot{x}_{13} =  a_1 \, x_{10}  \, x_6 - c_{6a} \, x_{13} - a_3 \, x_2  \, x_{13} + e_{2a} \, x_{14}, \\
        \dot{x}_{14} =  a_1 \, x_{11}  \, x_7 - e_{2a} \, k_v \, x_{14},                                     \\
        \dot{x}_{15} =  c_{2c} + c_{1c} \, x_7 - c_{3c} \, x_{15}
    \end{cases}
\end{equation}
and the outputs,~\eqref{nfkb:outputs}:
% \vspace{-2ex}
\begin{equation}
    \begin{cases}
        y_1 = x_2, \qquad y_2 = x_{10} + x_{13}, \qquad y_3 = x_9, \\
        y_4 = x_1 + x_2 + x_3, \qquad y_5 = x_7, \qquad y_6 = x_{12},
    \end{cases}
    \label{nfkb:outputs}
\end{equation}
We use the values of these parameters from \citep{tomasz_mathematical_2004} to reduce the number of target identifiability candidates: \(a_1, a_2, a_3, c_{1a}, c_{5a},c_{1c}, c_{3c}, c_{2c},%\]\[
c_1, c_2, c_3, c_4, e_{1a}, k_v\).
The output functions of~\eqref{nfkb} yields these weights ({not listed states} get weight of 1): \(c_5 \Rightarrow 3,~x_4, x_5, x_6, x_8, x_{11},  x_{14} \Rightarrow 2.\)

\subsection{Two SEIR epidemiological models}

The following two SEIR models were presented in~\citep[Examples~34 and~16]{massonis_structural_2020}.
Example 34 is presented in~\eqref{seir}, while example 16 is given by~\eqref{seir2}. We also provide our weight assignments.
\begin{align}
\begin{cases}                 \dot{S} = \Lambda - r \, \beta \,S \, I / N - \mu \, S,   \\
                 \dot{E} = \beta \, S \, I / N - \epsilon \, E - \mu \, e, \\
                 \dot{I} = \epsilon \, E - \gamma \, I - \mu \, I,         \\
                 \dot{R} = \gamma \, I - \mu \, R,                         \\
                 y = I + R.
             \end{cases}
         &
        \begin{tabular}{lcl}
            $E$      & $\implies$ & 2 \\
            $S$      & $\implies$ & 3 \\
            $\gamma$ & $\implies$ & 4
        \end{tabular}
  \label{seir}
\\
\begin{cases}
                 \dot{S} = -\beta \, S \, I,                  \\
                 \dot{E} = \beta \, S \, I - \epsilon \, E,   \\
                 \dot{I} = \epsilon \, E - (\rho + \mu) \, I, \\
                 \dot{R} = \rho \, I - d \, R,                \\
                 y = I + R
             \end{cases}
         &
        \begin{tabular}{lcl}
            $E$      & $\implies$ & 2 \\
            $S$      & $\implies$ & 3 \\
            $\beta$  & $\implies$ & 3 \\
            $\rho$   & $\implies$ & 3 \\
            $\gamma$ & $\implies$ & 4
        \end{tabular}
  \label{seir2}
\end{align}
The output functions for both examples are structurally similar. They are different from those in the original paper to increase the computational
difficulty for SIAN's Gr\"obner basis routine.

\subsection{HPV models}
We considered two HPV models studied in~\citep{brouwer_transmission_2015}. The model itself is given by~\eqref{HPV} with indices \(i,j\in \{F, \,M\}\). We present the Gr\"obner basis computation timings for two cases of outputs given in equations~\eqref{hpv_gr4} and \eqref{hpv_gr1}. The outputs in~\eqref{hpv_gr4} result in the weight of 2 assigned to the following parameters and states provided in~\eqref{hpv4weights} where \(i,j \in \{M, F\}\). Everything else gets weight  1. 
The output collection from~\eqref{hpv_gr1} results in~\eqref{hpv_gr1_weights}.
{\small
\begin{align*}
     & \begin{cases}
           \dot{S}_i      & = \frac{\mu}{2} + \gamma^G_{i}  \, I^G_i + \gamma^O_{i}  \, I^O_i - S_i  \, \mu  - S_i  \, (\beta^{OO}_{ji}   (I^O_i + I^{OG}_i) \\ & + \beta^{GO}_{ji}  (I^G_i + I^{OG}_i)) - S_i  (\beta^{OG}_{ji}   (I^O_i + I^{OG}_i) + \\
                          & + \beta^{GG}_{ji}  (I^G_i + I^{OG}_i)),                                                                                          \\
           \dot{I}^O_i    & = S_i  \, (\beta^{OO}_{ji}  \, (I^O_i + I^{OG}_i) + \beta^{GO}_{ji}  \, (I^G_i + I^{OG}_i)) + \gamma^G_{i}  \, I^{OG}_i          \\ & - I^O_i  \, (\nu^{OG}_{M} + \gamma^O_{i} + \mu + \beta^{OG}_{ji}  \, (I^O_i + I^{OG}_i) + \\
                          & + \beta^{GG}_{ji}  \, (I^G_i + I^{OG}_i)),                                                                                       \\
           \dot{I}^G_i    & = S_i  \, (\beta^{OG}_{ji}  \, (I^O_i + I^{OG}_i) + \beta^{GG}_{ji}  \, (I^G_i + I^{OG}_i)) + \gamma^O_{i}  \, I^{OG}_i          \\ & - I^G_i  \, (\nu^{GO}_{M} + \gamma^G_{i} + \mu + \beta^{OO}_{ji}  \, (I^O_i + I^{OG}_i) +  \\
                          & + \beta^{GO}_{ji}  \, (I^G_i + I^{OG}_i)),                                                                                       \\
           \dot{I}^{OG}_i & = I^O_i  \, (\nu^{OG}_{M} + \beta^{OG}_{ji}  \, (I^O_i + I^{OG}_i) + \beta^{GG}_{ji}  \, (I^G_i + I^{OG}_i))                     \\ &+ I^G_i  \, (\nu^{GO}_{M} + \beta^{OO}_{ji}  \, (I^O_i + I^{OG}_i) + \beta^{GO}_{ji}  \, (I^G_i + I^{OG}_i)) \\
                          & -  I^{OG}_i  \, (\gamma^O_{i} + \gamma^G_{i} + \mu),
       \end{cases} \numberthis \label{HPV} \\
     & \text{Output set 1:}
    \begin{cases}
        y_1  = I^G_M + I^{OG}_M,\ \ \ y_2 = I^O_M + I^{OG}_M, \\
        y_3  = I^{OG}_M + I^{OG}_F
    \end{cases}                        \numberthis \label{hpv_gr1}                                                                                                                                                                                                        \\
     & \text{Output set 2:}
    \begin{cases}
        y_1 = I^G_M + I^{OG}_M,\ \ \ y_2 = I^O_M + I^{OG}_M, \\
        y_3 = I^G_F + I^{OG}_F,\ \ \ y_4 = I^O_F + I^{OG}_F
    \end{cases}    \numberthis \label{hpv_gr4}                                                                                                                                                                                                                  \\
    \\
     & \text{Output set 1 weights: }\begin{cases}
                                        I^G_F, I^O_F, I^{OG}_F, S_M \Rightarrow 2            \\
                                        S_F, \gamma^G_F, \gamma^O_F, \nu^{GO}_F, \nu^{OG}_F, \\
                                        \beta^{GG}_{FM}, \beta^{GO}_{MF}, \beta^{OG}_{MF}, \beta^{OO}_{MF} \Rightarrow 3
                                    \end{cases} \numberthis \label{hpv_gr1_weights}
    \\
     & \text{Output set 2 weights: }\begin{cases}
                                        S_i, \gamma^G_i, \gamma^O_i,  \mu, \nu^{GO}_i, \nu^{OG}_i, \beta^{GG}_{ij}, \beta^{GG}_{ji}, \beta^{GO}_{ij}, \\
                                        \beta^{GO}_{ji}, \beta^{OG}_{ij}, \beta^{OG}_{ji},  \beta^{OO}_{ij}, \beta^{OO}_{ji}
                                    \end{cases} \numberthis \label{hpv4weights}
\end{align*}
}

\vspace{-0.05in}
\subsection{Example with slowdown: a SIR-model}\label{sec:slowdown}
In~\eqref{eq:siraqj} from \cite[Table 1, ID 26]{massonis_structural_2020}, we present an example in which the weight assignment generated by our algorithm \emph{increases} the running time of F4. In {\sc Maple}, we observed an increase in CPU time from around 12 to 50 minutes {while} memory usage slightly decreases from 11.5 to 10.8 GB. In Magma, this system shows a larger increase in memory from 5.6 to 18.8 GB with an increase in CPU time from around 7 to 32 minutes.
\begin{equation}
    \label{eq:siraqj}
    \begin{cases}
        \dot{S} = b\,N - S\,(I\,\lambda + \lambda\,Q\,\epsilon_a\,\epsilon_q + \lambda\,\epsilon_a\,A + \lambda\,\epsilon_j\,J + d+1),             \\
        \dot{I} = k_1\,A - (g_1 + \mu_2 + d_2)\,I,                                                                                                 \\
        \dot{R} = g_1\,In + g_2\,J - d_3\,R,                                                                                                       \\
        \dot{A} = S\,(I\,\lambda + \lambda\,Q\,\epsilon_a\,\epsilon_q + \lambda\,\epsilon_a\,A + \lambda\,\epsilon_j\,J) - (k_1 + \mu_1 + d_4)\,A, \\
        \dot{Q} = \mu_1\,A - (k_2 + d_5)\,Q,                                                                                                       \\
        \dot{J} = k_2\,Q + \mu_2\,I - (g_2 + d_6)\,J,                                                                                              \\
        y_1 = Q,                                                                                                                                   \\
        y_2 = J                                                                                                                                    \\
    \end{cases}
\end{equation}

\begin{table}[hbt!]
    {
        
        \centering
        {\tiny
            \begin{tabular}{||c|c|c||c|c|c||c|c|c||c|c||}
                \hline
                \multicolumn{3}{||c||}{Model information} & \multicolumn{3}{c||}{Time (min)} & \multicolumn{3}{c||}{Memory (GB)} & \multicolumn{2}{c||}{Primes}                                                                                         \\
                \hline
                Model                                     & num.                             & num.                              & old SIAN                     & differential & our final  & speedup & old SIAN & differential & our final & reduction \\
                name                                      & polys.                           & vars.                             & ordering                     & degrevlex    & ordering   &         & ordering & degrevlex    & ordering  &           \\\hline\hline
                HPV,                                      &                                  &                                   &                              &              &            &         &          &              &           &           \\
                \eqref{HPV}, \eqref{hpv_gr4}                  & 79                               & 75                                & N/A                          & N/A          & N/A        & N/A     & N/A      & N/A          & N/A       & N/A       \\\hline
                COVID Model 2,                            &                                  &                                   &                              &              &            &         &          &              &           &           \\
                \eqref{ssaair}                              & 49                               & 48                                & N/A                          & 3762.7       & \(\infty\) & N/A     & 23.5     & \(\infty\)   & N/A       & 84        \\\hline
                Pharmacokinetics,                         &                                  &                                   &                              &              &            &         &          &              &           &           \\
                \eqref{pharm}                               & 48                               & 47                                & N/A                          & 102.5        & \(\infty\) & N/A     & 8.0      & \(\infty\)   & N/A       & 212       \\\hline
                HPV,                                      &                                  &                                   &                              &              &            &         &          &              &           &           \\
                \eqref{HPV}, \eqref{hpv_gr1}                  & 97                               & 92                                & N/A                          & 179.0        & \(\infty\) & N/A     & 4.0      & \(\infty\)   & N/A       & 82        \\\hline
                Goodwin Oscillator                        &                                  &                                   &                              &              &            &         &          &              &           &           \\
                \eqref{goodwin}                             & 42                               & 43                                & 148.6                        & 7.8          & 19.0       & 11.0    & 0.8      & 13.7         & 103       & 103       \\\hline
                SEIR-1,                                   &                                  &                                   &                              &              &            &         &          &              &           &           \\
                \eqref{seir}                                & 44                               & 45                                & 10.2                         & 0.8          & 12.6       & 3.3     & 0.1      & 24.7         & 68        & 68        \\\hline
                COVID Model 1,                            &                                  &                                   &                              &              &            &         &          &              &           &           \\
                \eqref{seiqrdc}                             & 51                               & 50                                & 1346.9                       & 5.0          & 173.9      & 15.7    & 0.3      & 45.7         & 120       & 120       \\\hline
                NF-\(\kappa\)B,                           &                                  &                                   &                              &              &            &         &          &              &           &           \\
                \eqref{nfkb}                                & 120                              & 109                               & 40.6                         & 13.7         & 3.0        & 6.3     & 3.2      & 1.9          & 88        & 89        \\\hline
                SEIRP,                                    &                                  &                                   &                              &              &            &         &          &              &           &           \\
                \eqref{seirp}                               & 50                               & 42                                & 8.3                          & 3.5          & 2.4        & 1.7     & 0.3      & 6.3          & 32        & 32        \\\hline
                SEIR-2,                                   &                                  &                                   &                              &              &            &         &          &              &           &           \\
                \eqref{seir2}                               & 44                               & 43                                & 2.9                          & 1.7          & 1.8        & 0.8     & 0.2      & 4.6          & 38        & 38        \\\hline
            \end{tabular}
        }
        \caption{Weighted ordering applied to the Gr\"obner basis computation step of SIAN with zero characteristic  using {\sc Maple} 2021.2. ``N/A'' stands for the {\sc Maple} error ``\texttt{Error, (in Groebner:-F4:-GroebnerBasis) numeric exception:~division by zero}''  without a clear direct cause.}
        \label{benchmarks_maple_zero_char}
    }
\end{table}
\begin{table}[hbt!]
    {
       
        \centering
        {\tiny
            \begin{tabular}{||c|c|c||c|c|c||c|c|c||c|c||}
                \hline
                \multicolumn{3}{||c||}{Model information} & \multicolumn{3}{c||}{Time (min)} & \multicolumn{3}{c||}{Memory (GB)} & \multicolumn{2}{c||}{Primes}                                                                                        \\
                \hline
                Model                                     & num.                             & num.                              & old SIAN                     & differential & our final & speedup & old SIAN & differential & our final & reduction \\
                name                                      & polys.                           & vars.                             & ordering                     & degrevlex    & ordering  &         & ordering & degrevlex    & ordering  &           \\\hline\hline
                HPV,                                      &                                  &                                   &                              &              &           &         &          &              &           &           \\
                \eqref{HPV}, \eqref{hpv_gr4}                  & 79                               & 75                                & 31.2                         & 85.2         & 0.4       & 2.8     & 2.7      & 1.0          & 8         & 74        \\\hline
                COVID Model 2,                            &                                  &                                   &                              &              &           &         &          &              &           &           \\
                \eqref{ssaair}                              & 49                               & 48                                & 20722.2                      & 6784.6       & 3.1       & 38.6    & 24.1     & 1.6          & 36        & 93        \\\hline
                Pharmacokinetics,                         &                                  &                                   &                              &              &           &         &          &              &           &           \\
                \eqref{pharm}                               & 48                               & 47                                & 1181.6                       & 202.7        & 5.8       & 9.4     & 10.8     & 0.9          & 17        & 275       \\\hline
                HPV,                                      &                                  &                                   &                              &              &           &         &          &              &           &           \\
                \eqref{HPV}, \eqref{hpv_gr1}                  & 97                               & 92                                & 1229.2                       & 457.4        & 2.7       & 10.6    & 18.8     & 0.6          & 167       & 185       \\\hline
                Goodwin Oscillator                        &                                  &                                   &                              &              &           &         &          &              &           &           \\
                \eqref{goodwin}                             & 42                               & 43                                & 112.5                        & 3.5          & 31.8      & 3.1     & 0.5      & 6.0          & 109       & 169       \\\hline
                SEIR-1,                                   &                                  &                                   &                              &              &           &         &          &              &           &           \\
                \eqref{seir}                                & 44                               & 45                                & 18.5                         & 0.3          & 73.2      & 2.1     & 0.2      & 9.9          & 84        & 87        \\\hline
                COVID Model 1,                            &                                  &                                   &                              &              &           &         &          &              &           &           \\
                \eqref{seiqrdc}                             & 51                               & 50                                & 7626.9                       & 3.1          & 2468.9    & 8.8     & 1.8      & 4.9          & 237       & 260       \\\hline
                NF-\(\kappa\)B,                           &                                  &                                   &                              &              &           &         &          &              &           &           \\
                \eqref{nfkb}                                & 120                              & 109                               & 78.7                         & 12.5         & 6.3       & 2.8     & 0.7      & 4.8          & 9         & 107       \\\hline
                SEIRP,                                    &                                  &                                   &                              &              &           &         &          &              &           &           \\
                \eqref{seirp}                               & 50                               & 42                                & 41.5                         & 3.3          & 12.6      & 1.3     & 1.6      & 0.8          & 62        & 41        \\\hline
                SEIR-2,                                   &                                  &                                   &                              &              &           &         &          &              &           &           \\
                \eqref{seir2}                               & 44                               & 43                                & 7.6                          & 1.1          & 6.9       & 1.4     & 0.7      & 1.7          & 80        & 78        \\\hline
            \end{tabular}
        }
        \caption{Weighted ordering applied to the Gr\"obner basis computation step of SIAN with zero characteristic using Magma V2.26-8.}
        \label{benchmarks_magma_zero_char}
    }
\end{table}

\section{Inverted weights}
The weight assignment we discussed above is not unique. In fact, we can even find an alternative assignment given the same weight generation procedure as described earlier. Instead of simply using the rule of higher level, as in \eqref{eq:level}, we can generate the following assignment:
\begin{equation*}
    \label{eq:inv_weights}
    \mathrm{Weight(x)} := M - \mathrm{Level}(x) + 1,
\end{equation*}
where \(M = \max\limits_x{\mathrm{Level}(x)}\) is a maximal possible level of each state. With this strategy, we can also see improvement. In fact, for certain systems, such as~\eqref{goodwin}, this assignment is more beneficial in reducing the runtime. At the same time, in case of~\eqref{HPV},~\eqref{hpv_gr4}, we observe a similar error message in {\sc Maple} as if weights were not present. The results of this new assignment are presented in Table~\ref{benchmarks_maple_inv}.
It shows that there is still room for improvement in finding a weight assignment rule.

One would expect that many of the same phenomena from the previous weight assignment would happen with these models as well.
In particular, it should attempt to force the variables of \emph{low} level to appear in the leading monomials of the corresponding equations, thus making the pairs involving these equations to be of higher formal degree, which would, once again, avoid selecting too many pairs at once and steer the computation towards first working with the variables and equations at a \emph{high} level only.
This heuristic is similar to how one would solve a ``simple'' ODE system by integration. 
In particular, if one would try to solve the equations  
\begin{align}
& x_2' = a  \label{eq:invert_eq1}      \\
     & x_1' = x_2 \label{eq:invert_eq2}                        \\
     & y = x_1 \label{eq:invert_eq3} ,    
\end{align}
one would first solve for $x_2$ by integrating the constant $a$ in~\eqref{eq:invert_eq1} to get $x_2 = at + x_2(0)$, then would integrate $x_1  $ in~\eqref{eq:invert_eq2} to get $x_1 = at^2 + x_2(0)t + x_1(0)$, and finally after substituting for $y$ in~\eqref{eq:invert_eq3} we get $y(t) = at^2 + x_2(0)t + x_1(0) $. 
The inverted weight assignment heuristic models the way a human would naturally solve such an ODE system. 
While we expect the regular weight assignment to be useful more often, we also expect there to be ODE systems that have a nice structure in the equations coming from higher derivatives that the inverted weights select first before other equations.
\begin{table}[t]
    \renewcommand*{\arraystretch}{0.9}
    \centering
    {\tiny
        \begin{tabular}{||c|c|c||c|c|c||c|c|c||}
            \hline
            \multicolumn{3}{||c||}{Model information} & \multicolumn{3}{c||}{Time (min)} & \multicolumn{3}{c||}{Memory (GB)}                                                                                                                                                 \\
            \hline
            Model                                     & num.                             & num.                              & eq. \eqref{eq:default_order} & eq. \eqref{eq:default_order}, & speedup    & eq. \eqref{eq:default_order} & eq. \eqref{eq:default_order}, & reduction  \\
            name                                      & polys.                           & vars.                             & order                      & inv. weights                &            & order                      & inv. weights                &            \\\hline\hline
            HPV,                                      &                                  &                                   &                            &                             &            &                            &                             &            \\
            \eqref{HPV}, \eqref{hpv_gr4}                  & 79                               & 75                                & N/A                        & N/A                         & -          & N/A                        & N/A                         & -          \\\hline
            COVID Model 2,                            &                                  &                                   &                            &                             &            &                            &                             &            \\
            \eqref{ssaair}                              & 49                               & 48                                & N/A                        & 607.1                       & \(\infty\) & N/A                        & 33.6                        & \(\infty\) \\\hline
            Pharmacokinetics,                         &                                  &                                   &                            &                             &            &                            &                             &            \\
            \eqref{pharm}                               & 48                               & 47                                & N/A                        & 127.0                       & \(\infty\) & N/A                        & 6.02                        & \(\infty\) \\\hline
            HPV,                                      &                                  &                                   &                            &                             &            &                            &                             &            \\
            \eqref{HPV}, \eqref{hpv_gr1}                  & 97                               & 92                                & N/A                        & 19.1                        & \(\infty\) & N/A                        & 3.4                         & \(\infty\) \\\hline
            Goodwin Oscillator                        &                                  &                                   &                            &                             &            &                            &                             &            \\
            \eqref{goodwin}                             & 42                               & 43                                & 29.8                       & 0.6                         & 72.6       & 10.6                       & 0.1                         & 21.7       \\\hline
            SEIR-1,                                   &                                  &                                   &                            &                             &            &                            &                             &            \\
            \eqref{seir}                                & 44                               & 45                                & 2.2                        & 0.23                        & 14.9       & 3.3                        & 0.1                         & 16.2       \\\hline
            COVID Model 1,                            &                                  &                                   &                            &                             &            &                            &                             &            \\
            \eqref{seiqrdc}                             & 51                               & 50                                & 321.9                      & 148.2                       & 2.2        & 15.2                       & 3.4                         & 4.4        \\\hline
            NF-\(\kappa\)B,                           &                                  &                                   &                            &                             &            &                            &                             &            \\
            \eqref{nfkb}                                & 120                              & 109                               & 7.1                        & 5.3                         & 1.3        & 6.1                        & 1.9                         & 3.2        \\\hline
            SEIRP,                                    &                                  &                                   &                            &                             &            &                            &                             &            \\
            \eqref{seirp}                               & 50                               & 42                                & 2.0                        & 4.5                         & 0.5        & 1.6                        & 0.8                         & 2.1        \\\hline
            SEIR-2,                                   &                                  &                                   &                            &                             &            &                            &                             &            \\
            \eqref{seir2}                               & 44                               & 43                                & 0.8                        & 0.7                         & 1.2        & 0.7                        & 0.1                         & 6.7        \\\hline
        \end{tabular}
    }
    \caption{Results of applying the \emph{inverted} weighted ordering to only Gr\"obner basis computation step of SIAN with  characteristic \(p=11863279\) using {\sc Maple} 2021.2. ``N/A'' stands for the {\sc Maple} error 
        ``\texttt{Error, (in Groebner:-F4:-GroebnerBasis) numeric exception:~division by zero}'' without a clear direct cause.}
    \label{benchmarks_maple_inv}
\end{table}
\end{document}